\documentclass[12pt,english,prl, onecolumn, notitlepage, reprint]{revtex4-1}

\usepackage[T1]{fontenc}
\usepackage[latin9]{inputenc}
\usepackage{color}
\usepackage{bm}
\usepackage{amsmath}
\usepackage{amssymb}
\usepackage{graphicx}
\usepackage{textcomp}

\makeatletter
\usepackage{amsmath}
\usepackage{subfigure}
\usepackage{graphicx, mathtools}
\usepackage[normalem]{ulem}
\usepackage[colorlinks=true,linkcolor=MidnightBlue,urlcolor=MidnightBlue,citecolor=MidnightBlue,anchorcolor=MidnightBlue]{hyperref}
\usepackage[dvipsnames]{xcolor}
\usepackage{mathptmx, txfonts}
\usepackage{setspace} 
\usepackage[english]{babel}
\newcommand{\beginsupplement}{%
        \setcounter{table}{0}
        \renewcommand{\thetable}{S\arabic{table}}%
        \setcounter{figure}{0}
        \renewcommand{\thefigure}{S\arabic{figure}}%
        \setcounter{equation}{0}
        \renewcommand{\theequation}{S\arabic{equation}}%
     }

\makeatother

\usepackage{babel}
\begin{document}
\title{Flow-induced phase separation of active particles is controlled by boundary conditions}\vspace*{0.5cm}

\author{Shashi Thutupalli}
\email{shashi@ncbs.res.in}
\affiliation{Simons
Centre for the Study of Living Machines, National Centre for Biological
Sciences, GKVK Campus, Bellary Road, Bangalore 560065, India}
\affiliation{International Centre for Theoretical Sciences, Tata Institute
of Fundamental Research, Bangalore 560012, India}
\affiliation{Department of Mechanical and Aerospace Engineering, Princeton
University, Princeton, NJ 08544, USA}
\affiliation{Joseph Henry Laboratories of Physics, Princeton University,
Princeton, NJ 08544, USA}

\author{Delphine Geyer}
\affiliation{Department of Mechanical and Aerospace Engineering, Princeton
University, Princeton, NJ 08544, USA}

\author{Rajesh Singh}
\affiliation{The Institute of Mathematical Sciences-HBNI, CIT Campus, Chennai
600113, India}
\affiliation{DAMTP, Centre for Mathematical Sciences, University of Cambridge,
Wilberforce Road, Cambridge CB3 0WA, UK}

\author{Ronojoy Adhikari}
\affiliation{The Institute of Mathematical Sciences-HBNI, CIT Campus, Chennai
600113, India}
\affiliation{DAMTP, Centre for Mathematical Sciences, University of Cambridge,
Wilberforce Road, Cambridge CB3 0WA, UK}

\author{Howard A. Stone}
\affiliation{Department of Mechanical and Aerospace Engineering, Princeton
University, Princeton, NJ 08544, USA}

\begin{abstract}
Active particles, including swimming microorganisms, autophoretic
colloids and droplets, are known to self-organize into ordered structures at fluid-solid boundaries. The entrainment of particles in the attractive parts of
their spontaneous flows has been postulated as a possible mechanism
underlying this phenomenon. Here, combining experiments, theory and
numerical simulations, we demonstrate the validity of
this flow-induced ordering mechanism in a suspension of active emulsion
droplets. We show that the mechanism can be controlled, with a variety
of resultant ordered structures, by simply altering hydrodynamic
boundary conditions. Thus, for flow in Hele-Shaw cells, metastable
lines or stable traveling bands can be obtained by varying the cell
height. Similarly, for flow bounded by a plane, dynamic crystallites are formed. At a no-slip wall the crystallites are characterised by a continuous out-of-plane flux of particles that circulate and re-enter at the crystallite edges, thereby stabilising them. At an interface where the tangential stress vanishes the crystallites are strictly two-dimensional, with no out-of-plane flux. We rationalize these experimental results by calculating, in each case, the slow viscous flow produced by the droplets and the dissipative, long-ranged, many-body active forces and torques between them. The results of numerical simulations of motion under the action of the active forces and torques
are in excellent agreement with experiments. Our work elucidates the
mechanism of flow-induced phase separation (FIPS) in active fluids, particularly active colloidal suspensions,
and demonstrates its control by boundaries, suggesting new routes
to geometric and topological phenomena in active matter. 
\end{abstract}

\maketitle
{T}here are many instances, drawn from biological, physico-chemical
and technological contexts, in which microscopic particles produce
spontaneous flow in a viscous fluid. The energy necessary to maintain
this flow is supplied by a variety of mechanisms, of which there are a wide variety, at the interface
between the particles and the fluid. The ciliary layer in cells \cite{brennen1977},
the chemically reacting boundary layer in autophoretic colloids \cite{ebbens2010pursuit},
and the dissolution layer in auto-osmophoretic drops \cite{Thutupalli2011}
provide three distinct examples. In each case, the activity within
the layer drives the exterior fluid into motion which appears as if
it were a spontaneous fluid flow around the particles. It is possible,
though not necessary, for the particles to translate and/or rotate
in response to the spontaneous flow. Irrespective of the property
of self-propulsion and/or self-rotation, such active particles in a
suspension will each produce a spontaneous flow in which other particles
will be entrained. This mutual entrainment, if sufficiently strong,
can produce states of organization with no analogue in an equilibrium
suspension of passive particles. 

The above-mentioned mechanism has been conjectured~\cite{singh2016crystallization} to underlie the
spontaneous crystallization of Janus particles \cite{palacci2013living}
and fast moving bacteria \cite{petroff2015fast} at a plane wall.
However, a conclusive experimental demonstration of the validity of
this flow-induced phase separation (FIPS) mechanism is still lacking.
If bulk hydrodynamic flow is the principal cause of self-organization,
any alteration of the flow should manifest itself in altered states
of self-organization. The simplest way of altering the bulk flow,
keeping other experimental conditions constant, is to vary the hydrodynamic
conditions at the boundaries of the flow. If this produces correspondingly
distinct states of self-organization, both the FIPS mechanism and
the role of boundaries in controlling it are, thereby, established.
\begin{figure*}
\centering
\includegraphics[width=0.56\textwidth]{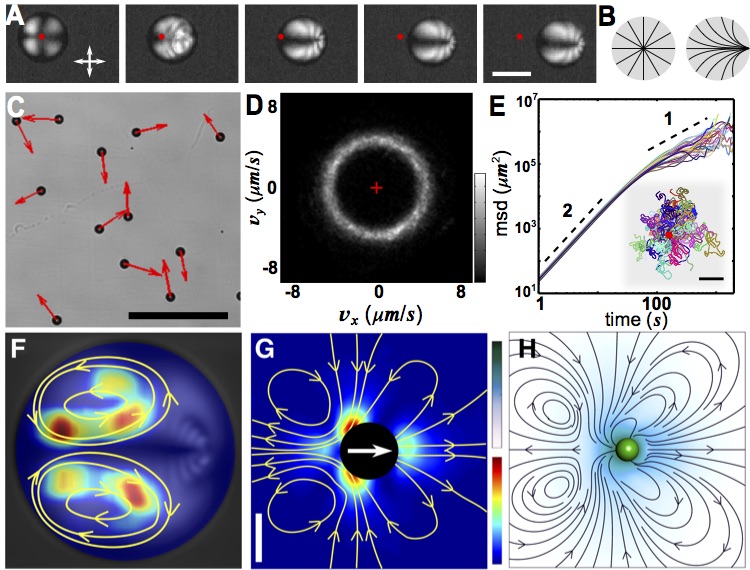}
\caption{Active droplets made with nematic liquid crystals. \textbf{A}. Cross
polarisation microscopy images showing spontaneous symmetry breaking
and propulsion of an active droplet. The red dot marks the initial position
of the active droplet. Each frame is $4$ seconds apart. Scale bar is $50$ \textmu m
. \textbf{B}. A sketch of the nematic director field (black lines)
inside the droplet due to homeotropic anchoring conditions at the
interface before (hedgehog defect in the center) and after the symmetry
breaking (escaped radial configuration). $\textbf{C}$. Active droplets
in a Hele-Shaw cell (height $50$~\textmu m, which is also the diameter
of the droplets; the lateral dimensions of the cell extend beyond
the field of view and are $4$ cm $\times$ $3$ cm. Scale bar is $500$~\textmu m. The red arrows indicate the instantaneous velocities of the droplets.
$\textbf{D}$. The probability distribution, using $\sim10^{7}$ measurements,
of the velocity vectors ($v_{x},v_{y}$) for individual droplets in
a Hele-Shaw geometry.  Colorbar represents the normalised probability. \textbf{E}. The mean squared displacements calculated
from the trajectories of a dilute suspension of active droplets (areal fraction
of droplets is $<0.5~\%$). Inset: A superposition of the trajectories
of the droplets with their point of origin aligned (marked by red
spot). Scale bar is $1$ mm. $\textbf{ F.}$ The rearrangement of the
director field inside a droplet swimming to the right is caused by
a spontaneous flow inside the droplet. \textbf{G}. Experimentally
measured external fluid flow for an active droplet (marked by black circle)
moving at a fixed velocity. $\textbf{H}$. The theoretical flow from
a truncated spherical harmonic expansion, with the expansion coefficient
estimated from the experimental flow in \textbf{G.} The droplet diameter
in \textbf{F} and \textbf{G} is $50$ \textmu m and the colorbars represent
the normalized logarithm of the local flow speed.\label{fig:1}}
\end{figure*}

Here, we use a suspension of active, self-propelled emulsion droplets
to investigate the role of hydrodynamics on their collective behavior.
The system has been used previously \cite{Thutupalli2011,Thutupalli2013,Herminghaus2014,Izri2014,Shani2014}
to provide insights into out-of-equilibrium phenomena relevant to self-organization in both natural \cite{Cavagna2014,Sokolov2007,petroff2015fast}
and synthetic active particle settings. The typical size $b\sim50$~\textmu m of the emulsion droplets and their typical self-propulsion
speed $v_{s}\sim5$~\textmu m/s implies that the Reynolds number $Re=v_{s}b/\nu\sim10^{-4}$
in a fluid with the kinematic viscosity $\nu$ of water. Fluid inertia
is negligible at such small $Re$ and the flow is described by the
Stokes equation. It is then possible to exploit the linearity of the
governing equations and use Green's function techniques to compute
the flow, the stress in the fluid, and the forces and torques between
the droplets for a variety of boundary conditions. We find distinct
states of aggregation as the boundary conditions are altered, which
both validates the conjectured hydrodynamic mechanism and opens up
a route to its manipulation and control. 

We should note the effect of steric confinement of active fluids on
self-organization has been studied thoroughly in the past \cite{woodhouse2012,Wioland2013,Bricard2015,Solon2015,Fily2014}.
This knowledge has been exploited in applications such as self-assembly,
meta-material synthesis and active fluid computation \cite{niu2017self,Souslov2017,Woodhouse2017}. However the resultant hydrodynamic effects of confinement can be distinct due to difference, for instance, in the slip properties of the boundaries even when the geometry of the confinement is identical.
Our work suggests an independent route to ordered states of active
matter by using boundaries to alter the \emph{hydrodynamic interactions in}, rather than the \emph{confinement of}, the system. With this remark
we now turn to our results. 

\begin{figure*}
\includegraphics[width=0.6\textwidth]{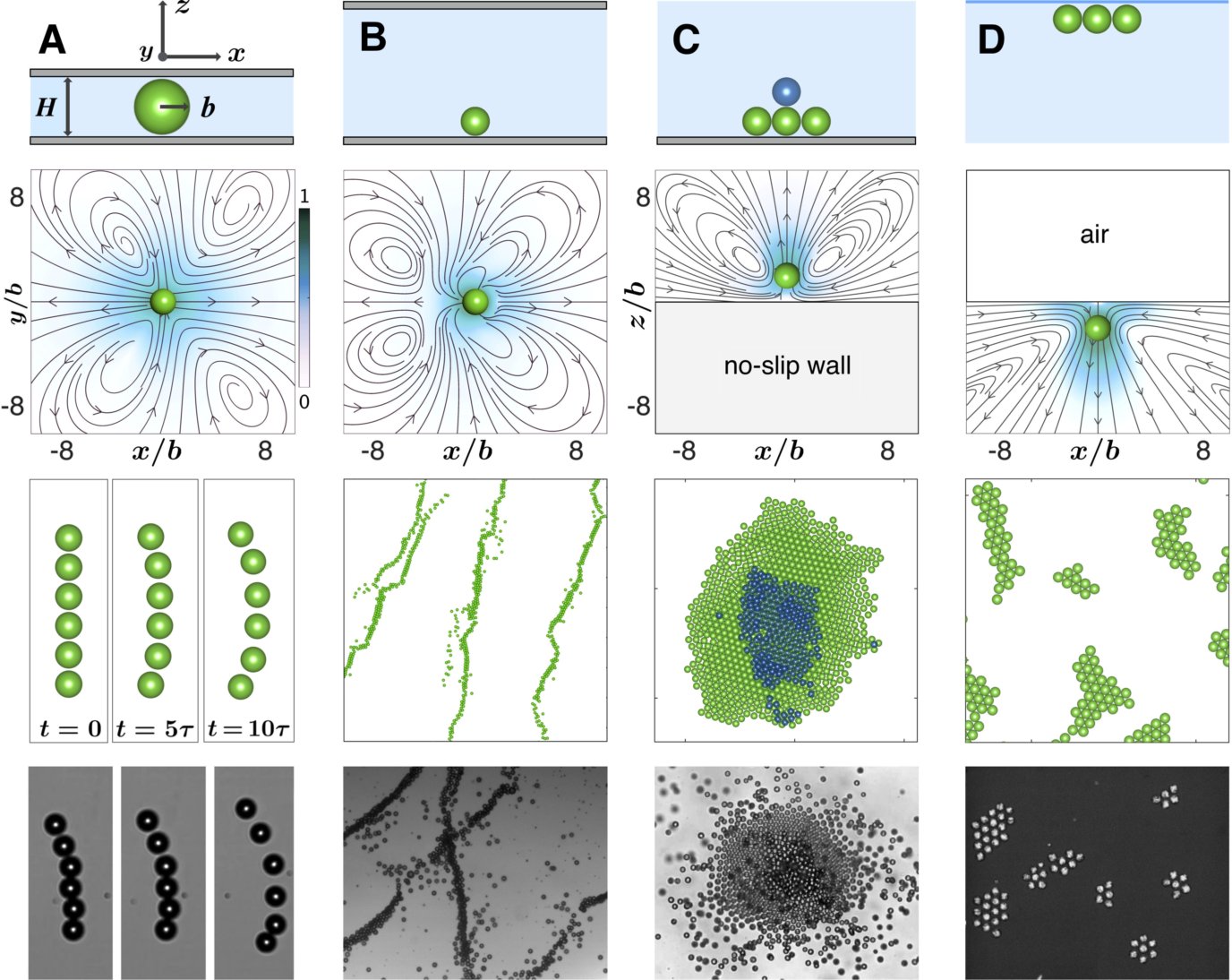}
\caption{The role of boundaries in determining the collective behaviour of
active droplets (particles). Top row: Schematic of confinement. Second row: The exterior
flow field produced by the active particles in each boundary condition considered.
Third and fourth (bottom) rows, respectively, contain snapshots from simulations
and experiments. Traveling lines of active droplets can be formed in a Hele-Shaw
cell. These lines are metastable i.e. they translate a few droplet diameters before breaking up, if the separation $H$ of the cell
is approximately equal to the droplet diameter (\textbf{A}, $H/b\sim2$), while these
lines are stable when the separation is a few droplet diameters (\textbf{B},
$H/b\sim8$). Aggregation of the droplets, leading to crystallization,
is observed at a plane wall (\textbf{C}, cell depth $H/b\sim400$)
and at the air-water interface (\textbf{D}, cell depth $H/b\sim400$).
At a plane wall, droplets are expunged from the crystalline core,
while the crystal is stabilized by the recirculation of the fluid
flow. At the air-water interface, on the other hand, there is no out-of-plane
motion. Here $\tau=b/v_{s}$ is the time in which the active droplet moves
a distance equal to its radius.\label{fig:2}}
\end{figure*}

\subsection*{Experimental system and theoretical model}
Our experimental system is an active emulsion of monodisperse droplets of
liquid crystal (5CB) in water whose source of activity is droplet
dissolution \cite{Herminghaus2014}. Surface tension gradients at
the interface of the droplet, sustained by the free energy of dissolution,
produce active hydrodynamic flows both within and external to the droplet, leading to droplet motion. While this self-propulsion does not
rely on the liquid crystallinity of the droplet, the nematic state
of 5CB within it enables the internal velocity field to be inferred
(Fig.~\ref{fig:1}\textbf{A} and Movie S1). The presence of surfactant at
the interface makes it energetically favorable for the rod-like 5CB
molecules to orient normal to it. Such a homeotropic boundary condition
on the director field enforces a point defect, which is located at
the center of the droplet (Fig.~\ref{fig:1}\textbf{B}) when surface tension
gradients (and hence fluid flow) are negligible. When observed in
polarised light between cross-polarizers, the droplet shows a four-lobed
pattern (left-most panel of Fig.~\ref{fig:1}\textbf{A}) reflecting the symmetric
orientation of the director field about the droplet center. When surface
tension gradients become appreciable, viscous stresses and fluid flow
are induced in the bulk, and so the nematic stress within the droplet
must be redistributed. The resulting reorientation of the director
causes a displacement of the point defect along the axis of droplet
motion and the centro-symmetric four-lobed pattern is distorted to
another with a reduced symmetry, now only about the propulsion axis
(Fig.~\ref{fig:1}\textbf{A, B}).

Each droplet propels in a random direction set by its own internal
spontaneously broken symmetry and these directions are distributed
isotropically (shown in Fig.~\ref{fig:1}\textbf{C} for the case of a quasi
two-dimensional Hele-Shaw cell). The droplet speed (Fig.~\ref{fig:1}\textbf{D})
is set by its size and the concentration of the surfactant in the
external phase \cite{Herminghaus2014}. In such a configuration, individual
droplets exhibit random, diffusive-like motion due to the fluctuations
in the self-propulsion mechanism and due to interactions with the
other droplets (Fig.~\ref{fig:1}\textbf{E}). The balance between viscous
and nematic stresses within the droplet tends to align the velocity
field with the director field (Fig.~\ref{fig:1}\textbf{F}) \cite{Prishchepa2005},
resulting in an asymmetry in the circulatory flow inside the droplets
(Movie S2), with a stagnation point close to the point defect. This
asymmetry also appears in the external flow generated by the droplets,
as can be seen in the velocity field (Fig.~\ref{fig:1}\textbf{G}) due
to a droplet moving with speed $v_{s}$. It is in this external flow
that other particles are entrained and is, therefore, the focus of our theoretical
model.

Our theoretical model for an active particle is a sphere of radius
$b$ with an active slip prescribed at its surface. As our primary
interest is in the external flow, we assume the internal flow to be
a rigid body motion. The fluid velocity on the boundary of the $i$-th
sphere, then, is
\begin{equation}
\boldsymbol{v}(\boldsymbol{R}_{i}+\boldsymbol{\rho}_{i})=\mathbf{V}_{i}+\boldsymbol{\Omega}_{i}\times\boldsymbol{\rho}_{i}+\boldsymbol{v}_{i}^{\mathcal{A}}(\boldsymbol{\rho}_{i}),\label{eq:slip-RBM-BC}
\end{equation}
where $\boldsymbol{R}_{i}$ is the center of the sphere, $\boldsymbol{\rho}_{i}$
is a point on its surface with respect to the center and $\mathbf{V}_{i}$
and $\boldsymbol{\Omega}_{i}$ are, respectively, its linear and angular
velocity. The active slip, $\boldsymbol{v}_{i}^{\mathcal{A}}(\boldsymbol{\rho}_{i})$,
is taken to be the most general surface vector field consistent with
incompressibility. Neither axial symmetry of the slip about the orientation
axis, $\boldsymbol{p}_{i}$, of the sphere nor flow purely tangential
to the interface is assumed. These assumptions distinguish our model~\cite{ghose2014irreducible,singh2015many,singh2016crystallization,singh2016generalized}
from that of the classical squirmer~\cite{lighthill1952,blake1971a}.
The translational and rotational velocities of the spheres are not
known \emph{apriori} but must be determined in terms of the slip velocities
from a balance of all forces and torques acting on them. 

\begin{figure*}
\includegraphics[width=0.6\textwidth]{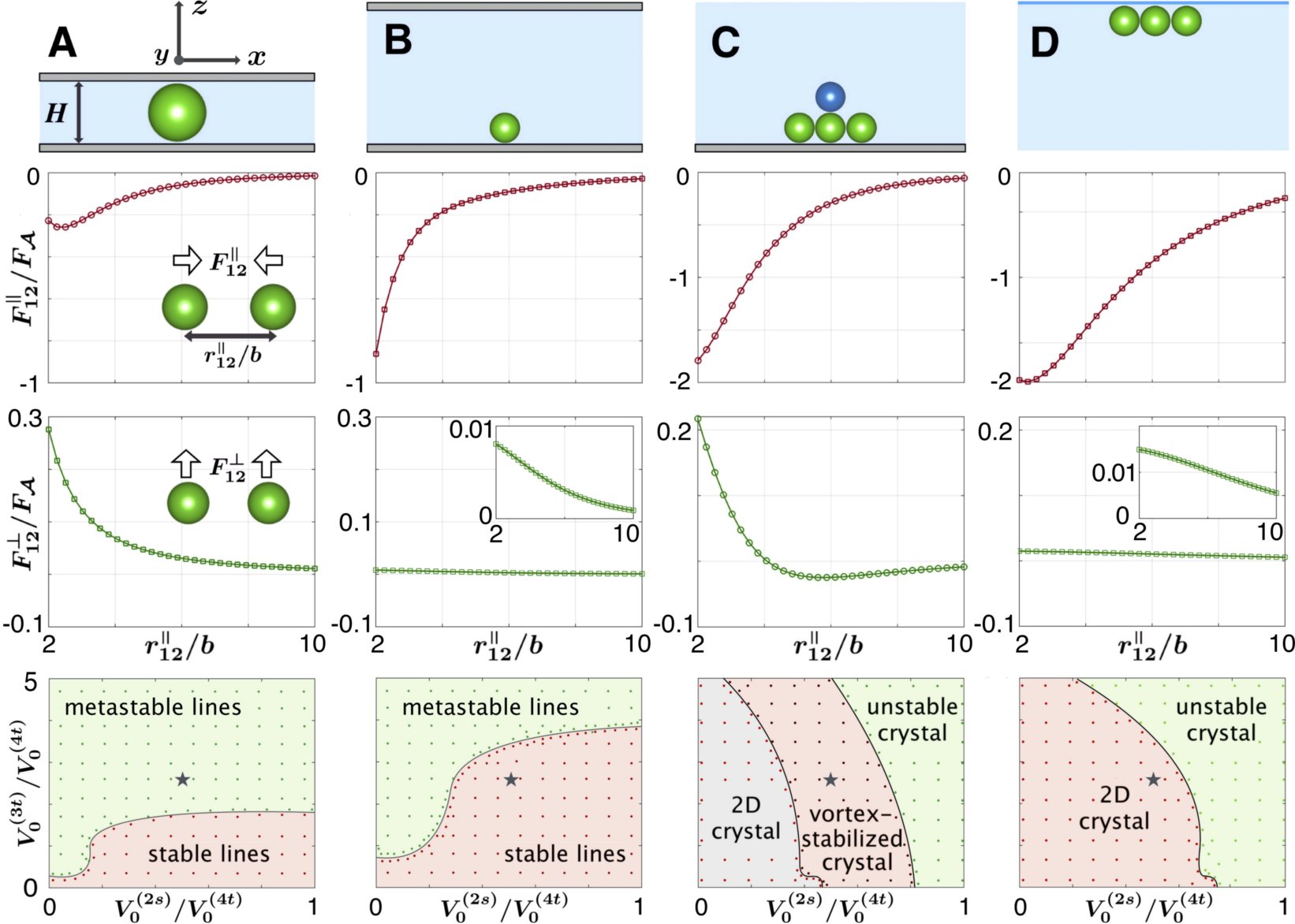}
\caption{Active forces on the active particles are modified by the presence of boundaries.
Second row: Attractive parallel forces lead to the formation of traveling
lines in a Hele-Shaw cell (\textbf{A} and \textbf{B}), and aggregation
in the plane of the wall (\textbf{C}) and the interface (\textbf{D}).
Third row: The perpendicular forces in a Hele-Shaw cell are an order
of magnitude larger for \textbf{A}, and accounts for the metastable
lines in this case. The perpendicular force at the plane wall is ten
times larger than the corresponding force at the interface and results
in the circulatory motion of active particles (see text for more details).
Insets show close-up and $F_{\mathcal{A}}=6\pi\eta bv_{s}.$ Bottom
row: State diagrams in terms of the strengths of the slip modes -
$V_{0}^{(2s)}$: symmetric dipole, $V_{0}^{(3t)}$: vector quadrupole,
and $V_{0}^{(4t)}$: degenerate octupole. Each dot represents a simulation,
while the star denotes the values used in the above rows.\label{fig:3}}
\end{figure*}

Slip induces flow in the exterior fluid and the stresses thus produced
act back on the sphere surface with a force per unit area $\boldsymbol{f}$.
Then $\mathbf{F}_{i}^{H}=\int\boldsymbol{f}\,d\text{S}_{i}$ and $\mathbf{T}_{i}^{H}=\int\boldsymbol{\rho}_{i}\mathbf{\times}\boldsymbol{f}\,d\text{S}_{i}$
are the net hydrodynamic force and torque on sphere $i$, which include
contributions from the usual Stokes drag, proportional to $\mathbf{V}_{i}$
and $\boldsymbol{\Omega}_{i}$, and from the active stresses, proportional
to $\boldsymbol{v}_{i}^{\mathcal{A}}$. The force per unit area is
computed from the solution of the Stokes equation satisfying Eq.~\ref{eq:slip-RBM-BC} at the sphere surfaces and the appropriate hydrodynamic boundary conditions
at the exterior boundaries. 

For our purpose, the boundary integral representation of Stokes flow
is most suited for obtaining the force per unit area, as the hydrodynamic
boundary conditions can be directly applied by choosing a suitable
Green's function. We employ this approach here and analytically solve
the resulting integral equations for the force per unit area, to leading
order in sphere separation, in a basis of tensorial spherical harmonics
with appropriate Green's functions. The forces and torques thus obtained
are inserted into force and torque balance equations that are integrated
numerically to obtain the translational and rotational motions of
the spheres (further details of the model and simulations are in
the \textit{SI Appendix}). 

The free parameters in our model, the sphere radius $b_{i}$ and the
slip velocity $\boldsymbol{v}_{i}^{\mathcal{A}}$, are determined
as follows. We set $b_{i}\sim50$~\textmu m which is the measured radius
of an undissoluted droplet. There is less than $1\%$ change to this
value during the course of the experiment. We use the exterior flow
of a single droplet to determine the slip, as the two are uniquely
related for any given hydrodynamic boundary condition. We parametrize
$\boldsymbol{v}_{i}^{\mathcal{A}}$ in terms of its first three tensorial
harmonic coefficients, as these fully account for the long-ranged
components of the exterior flow. We then estimate the coefficients
by minimizing the square deviation between the experimentally measured
flow and the three mode expansion. The exterior flow thus obtained (Fig.~\ref{fig:1}\textbf{H}) is in good agreement with the experimentally
measured flow (Fig.~\ref{fig:1}\textbf{G}). We note that the flow in the Hele-Shaw cell is used
for this estimation and comparison. 

To emphasise, we \emph{fit} the \emph{one-body} exterior flow to estimate
the active slip and then use it to \emph{predict} the \emph{many-body}
exterior flow and the many-body forces and torques for any given boundary
condition. We note that the three-mode expansion is not a limitation
of the theoretical model, which accommodates as many modes as may
be necessary to represent the exterior flow to the desired level of
accuracy. 

\begin{figure*}
\includegraphics[width=0.6\textwidth]{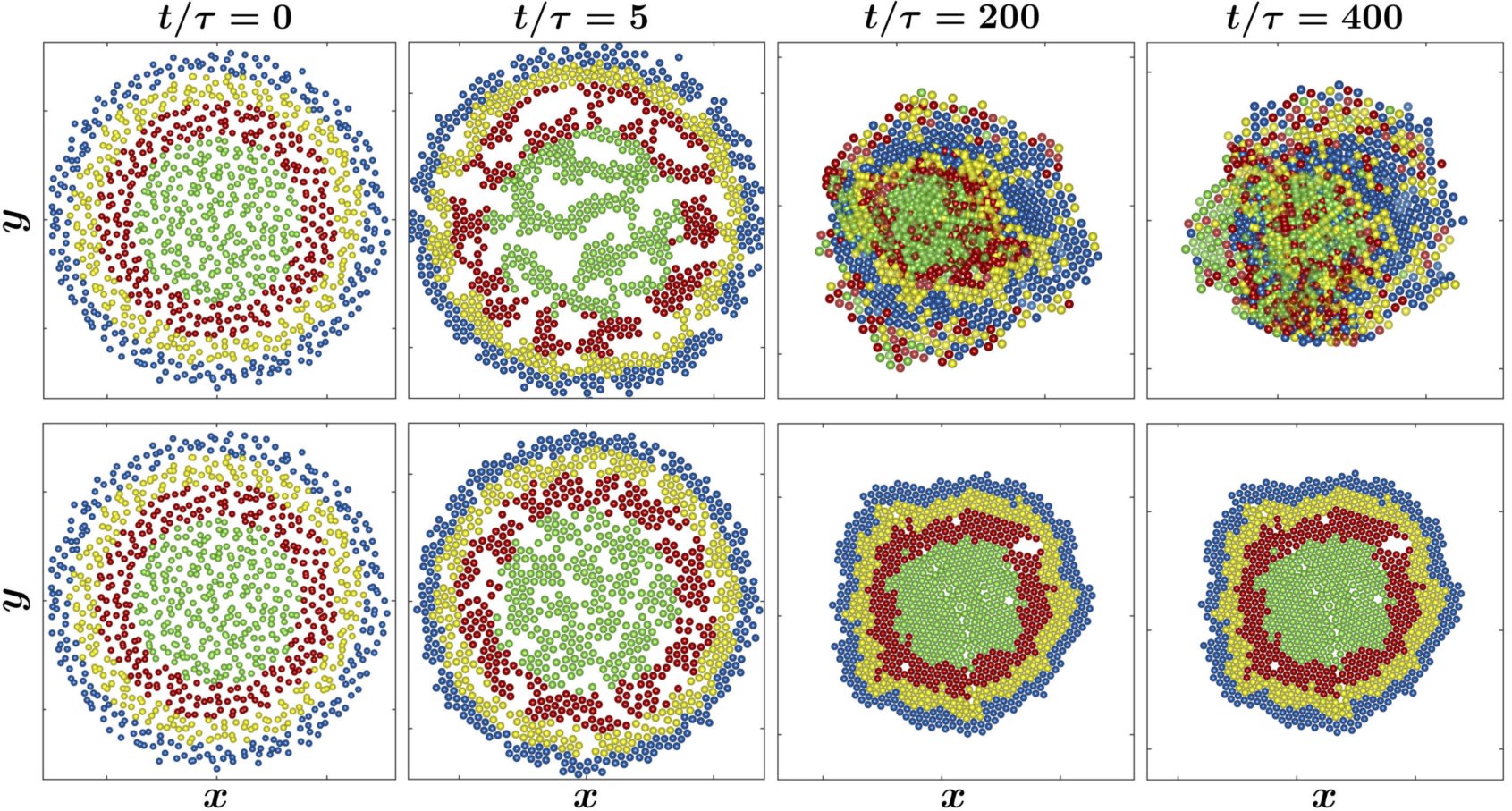}
\caption{Kinetics of aggregation of active particles at a plane wall (top row) and
at a plane interface (bottom row). The colors are used to indicate
and track the particles based on their initial positions. The instantaneous
snapshots show that there is a faster mixing of particles and exchange
of neighbors at the plane wall due to the circulatory flow streamlines
(Fig. \ref{fig:2}\textbf{C}) and higher values of perpendicular forces
(Fig. \ref{fig:3}\textbf{C}). \label{fig:4}}
\end{figure*}

\subsection*{Self-organization and boundary conditions}

We now present the main results on the correspondence between self-organization
of active particles and hydrodynamic boundary conditions. Our boundary
conditions are (i) a plane channel flow in a Hele-Shaw cell where
the channel width $H$ is approximately one particle diameter (Fig.~\ref{fig:2}\textbf{A}), (ii) a plane channel flow where the channel width
varies between several particle diameters ($H\sim6-10b$, Fig.~\ref{fig:2}\textbf{B}), (iii) flow bounded by a plane wall where the flow
vanishes (Fig.~\ref{fig:2}\textbf{C}) and (iv) flow bounded by a plane air-water interface
where the tangential stress vanishes (Fig.~\ref{fig:2}\textbf{D}). In every case, the emulsion parameters are kept unchanged;
the\emph{ only} change is in the exterior boundary conditions. 

Snapshots from the experiments (bottom row, Fig.~\ref{fig:2}) point to distinct signatures of the boundary conditions on the resultant self-organization of the droplets. In the Hele-Shaw cell, $H\sim2b$, droplets spontaneously
form metastable lines, which curve in their direction of motion and
eventually break up after traversing a few droplet diameters (Fig.~\ref{fig:2}\textbf{A} and Movie S3). Increasing the channel width,
$H\sim8b$, transforms these lines into surprisingly stable bands
that travel through each other even as they collide (Fig.~\ref{fig:2}\textbf{B} and Movie S4). In contrast, at a plane wall,
droplets form crystallites parallel to the plane. Droplets comprising the crystallite are constantly expelled from the center of the aggregate only to rejoin it at the edges. The recirculating flows ensure a balanced in and out-flux of the droplets and thereby maintain a constant mean droplet number within these aggregates (vortex-stabilised crystallites, Fig.~\ref{fig:2}C, Movie S5). When the plane
wall is replaced by a plane interface, the previous inflow and outflow 
is suppressed and the droplets form two-dimensional crystalline
aggregates. These aggregates are maintained in a steady state by a
continuous coagulation and fragmentation of the crystallites (Fig.~\ref{fig:2}\textbf{D}, Movie S6). 

A qualitative understanding of these states of self-organization is obtained from the \emph{one-body} external flow of the particle
in each of the four boundary conditions (corresponding
panels of Fig.~\ref{fig:2}). We emphasize, once again, that in this calculation the
active slip is \emph{estimated} from flow in the Hele-Shaw cell but
used to \emph{predict} flow for the three remaining boundary conditions.
Operationally, the latter only requires the use of the appropriate
Green's function. In the Hele-Shaw cell, the net flow is parallel to the walls and has an inflowing component
perpendicular to the direction of motion. Entrainment in this inflow
leads to the formation of metastable lines and stable bands. At both
the plane wall and the plane interface, the flow has a cylindrical
symmetry when the propulsion axis is perpendicular to the plane. In
the first case, the flow has a strong circulation in which entrained
particles are drawn inwards along the plane but then expelled normal
to it. In the second case, the circulatory component is comparatively
weak and entrained particles are primarily drawn inwards. Entrainment
in this flow leads to the formation, respectively, of vortex-stabilised crystallites and of coagulating and fragmenting two-dimensional
crystallites. The qualitative
agreement of the simulations (third row, Fig.~\ref{fig:2}) obtained from the numerical integration of the force and torque balance equations with experiment (bottom row, Fig.~\ref{fig:2}) is excellent. This agreement between experiment
and theory that disregards the internal flow confirms, \emph{a posteriori},
our hypothesis that entrainment in the external flow is primarily
responsible for self-organization. 

A quantitative understanding of the states of self-organization requires
an accurate estimate of the forces and torques between particles. We calculate the components of the active pair force
parallel and perpendicular to their separation vector, as a function
of separation distance, for each of the four boundary conditions considered (second and third rows, Fig.~\ref{fig:3}).
In each of the cases, the component of the active force parallel to
the separation vector is negative and it is this attractive component
of the active force that leads to aggregation. However, there is a
considerable variation in the component of the active force perpendicular
to the separation vector, and it is the sensitivity of this component
to boundary conditions that accounts for the variety of the aggregated
states. In the Hele-Shaw cell, the perpendicular force is positive
which, in our sign convention, means that it is directed along the
direction of motion. Since to leading order, hydrodynamic forces are
pair-wise additive, this implies that the net force on particles at
the center of a moving line are greater than those at the edges. Therefore,
they tend to move faster, creating a curvature of the line and eventually
its break up. In contrast, the perpendicular force in the plane channel
is an order of magnitude smaller (as shown in the corresponding inset
of Fig.~\ref{fig:3}\textbf{B} for clarity) and the break up mechanism
has a negligible contribution, which gives the traveling bands their
surprisingly stability. At a no-slip wall (Fig.\ \ref{fig:3}\textbf{C}),
the perpendicular force is negative at large distances but positive
at short distances. This drives particles into the wall when they are
well-separated but away from the wall when the are close by. Combined
with the parallel component of the flow, which is always attractive,
this leads to the expulsion and re-circulation of particles in the
crystalline aggregate. In contrast, the perpendicular force at an
interface, (Fig.\ \ref{fig:3}\textbf{D}) is positive, but an order
of magnitude smaller (again, shown in the inset for clarity) and the
dominant motion is due to the attractive component of the parallel
force. Thus, expulsion is suppressed and the result is the formation
of two-dimensional crystallites. Similar estimates for the torque
provide an understanding of the orientational dynamics which is relatively
unimportant in our case as Brownian re-orientation is negligible and
the hydrodynamic torques are one power of separation smaller than
the corresponding forces. 

Since each irreducible mode of the slip is independent of the others
and produces flow of distinct multipolar symmetry, it is possible
to isolate the effect of each mode on the self-organization. The modes
are labelled by an angular momentum index $l=0,1,2,\ldots$ and a
spin index $\sigma=s,a,t$ corresponding to the symmetric, antisymmetric
and pure trace irreducible components of each mode. Using the abbreviation
$l\sigma$ to indicate a mode with angular momentum index $l$ and
spin index $\sigma$, the $3t$ mode produces self-propulsion (and
a degenerate quadrupolar flow) while the $2s$ mode produces inflow
and outflow along mutually perpendicular axes (and a dipolar flow).
The latter is the stresslet mode and its instantaneous sign determines
if the fluid is being expelled (``pusher'') or ingested (``puller'')
along the propulsion axis. We are then able to construct a state-diagram
in the $V_{0}^{(2s)}-V_{0}^{(3t)}$ plane that demarcates regions
of stability of the principal states of aggregation found for each
of the four boundary conditions studied (Fig.~\ref{fig:3}, bottom row). In Hele-Shaw flow, the stresslet mode
promotes stability but in flows bounded by a plane wall, the perpendicular
component of the force is enhanced in proportion to the magnitude of the stresslet,
which leads from two-dimensional crystallites to vortex-stabilised crystallites and finally to instabilities. Thus, by suitably choosing
the strength of the stresslet mode, it is possible to select either
a state of two-dimensional crystals or vortex-stabilised crystals.
For flow bounded by an interface, an enhanced stresslet mode does
not lead to explusion of the droplets out-of-plane but directly to an instability (unstable crystallites). 

Our results above provide convincing evidence of the mechanism of
hydrodynamic entrainment in the spontaneous exterior flow as the dominant
mechanism for self-organization in active particles. The active parts
of the hydrodynamic forces and torques that result from this entrainment
provide both a qualitative and quantitative explanation of the states
of self-organization. These hydrodynamic forces and torques depend
on the distance between the particles, on their orientation, and the
magnitudes of the modes of the slip on each active particle. It can
be directly verified from the explicit forms of the forces and torques
that they cannot be obtained as gradients of potentials \cite{niu2017self}.
Further, their dependence on the modes of the slip velocity indicates
that they have odd parity under time reversal, signaling their explicitly
dissipative nature \cite{singh2016generalized,singh2016crystallization}.
Thus, we have a novel situation in which long-range, \emph{dissipative}
forces and torques promote self-organization. States of self-organization
maintained by entropy production were studied in the past by the Brussels
school and were given the name \emph{dissipative structures }\textit{\emph{\cite{kondepudi2014modern}}}\emph{.
}Self-organization in active particles, as shown here, appears to be an example of
a dissipative structure but one in which the dissipative mechanism
and the resultant forces and torques are unambiguously identified.

\subsection*{Kinetics of flow-induced phase separation}
The self-organization presented above can be viewed, from the point
of view of statistical physics, as a phase separation phenomenon driven
by dissipative forces, rather than the usual conservative forces derived
from a potential. The study of such flow-induced phase
separation (FIPS) presents several challenges, one of which is to
determine a quantity that can serve as the analogue of a thermodynamic
potential. It appears from recent theoretical work that large-deviation results
for the stationary distribution of Markov processes violating detailed
balance may provide a tractable route for answering questions of stability
\cite{touchette2009large}. Here we draw attention to the fact that
kinetic routes to similar flow-induced phase-separated states may
vary depending on the boundary condition. This is borne out in the differences between the kinetics of phase separation at
a plane wall and at a plane interface (Fig.~\ref{fig:4}
and Movie S7). At a plane wall, there is
both an enhanced mixing within the crystal plane and an exchange
of neighbors due to the closed streamlines (Fig.~\ref{fig:2}\textbf{C})
and higher values of perpendicular forces (Fig.~\ref{fig:3}\textbf{C})
compared to the plane interface. In a biological context, such hydrodynamic
bound structures and associated kinetics could influence the encounter
rate of individuals in aggregates \cite{Drescher2009,petroff2015fast}. 

\subsection*{Discussion}
The flow-induced phase separation (FIPS) mechanism established here
and conjectured earlier \cite{singh2016crystallization} has several
distinguishing features that bear pointing out and contrasting with
other mechanisms. First, self-propulsion and/or self-rotation are
\emph{not} necessary for its operation; it is only necessary that
the particles produce a long-ranged exterior hydrodynamic flow. The
experimentally observed states of self-organization persist even when
the self-propulsion parameter $V_{0}^{(3t)}$ vanishes (state diagrams, bottom row of Fig. \ref{fig:3}). Second, in
the absence of thermal fluctuations, the suspension is mechanically
unstable to aggregation at any positive value of the density and any
finite amount of activity, however small. It is plausible that in
the presence of thermal fluctuations, finite values of density and
activity are required to overwhelm the loss in entropy due to aggregation.
A careful study of aggregation at different temperatures, densities
and activities is needed to establish this quantitatively. Third,
FIPS includes as a special case, aggregation in driven colloidal systems
which are nonetheless force and torque free~\cite{Negi2005,Sapozhnikov2003}.
Fourth, the long-lived, stable traveling bands that we have shown
here are qualitatively different as compared to other dynamic behavior
seen in active systems \cite{niu2017self,Schaller2010,ohta2014traveling}
or in the emergence of orientational order in flocking models \cite{Solon2015a}.
In spite of the similarity in the aggregated states, the hydrodynamic
mechanism identified here is distinct. 
Fifth, our work provides an understanding of phase-separation in active systems that is complementary to motility-induced phase separation (MIPS) \cite{cates2015}. The latter has a kinematic character, in which the flux of particles is a prescribed functional of density, reflecting the tendency of active particles to slow down or speed up in regions of, respectively, higher or lower density. The physics underlying this tendency is left implicit, and may plausibly be attributed to non-hydrodynamic interactions, like contact, compression, or jamming. In contrast, FIPS has an explicitly dynamic character, as forces and torques of hydrodynamic origin are identified in causing the aggregation. It is entirely conceivable that a state of aggregation like  hexagonal crystallites can be formed from either of these mechanisms. On the other hand, vortex-stabilised crystallites appear difficult to explain within MIPS whereas they appear naturally in FIPS. Since FIPS needs a wall with no slip or an interface with vanishing tangential stress, aggregate structures that remain stable away from such boundaries would point to a mechanism such as MIPS as the source of stability.
Finally, though our experiments
are performed with droplets, the theory and simulation correspond
to arbitrary active particles and our flow boundary conditions are
generic to many natural and engineered settings. Together, these underscore
the importance of our findings to a wide class of active fluids, in particular active colloids, and
to the study and control of geometric and topological phenomena in active matter. 

\subsection*{Acknowledgements} We thank T. Tlusty, M.E. Cates and R. E. Goldstein for discussions. S.T. acknowledges the Human Frontier Science Program (Cross Disciplinary Fellowship) for funding. R.A. thanks the IUSSTF for supporting a sabbatical visit to Princeton University. R.S and R.A. acknowledge the Institute of Mathematical Sciences for computing resources on the Nandadevi clusters.

\beginsupplement
\section*{Supplementary Information (SI)}
\subsection*{Experimental methods}
Our experimental system is comprised of thousands of monodisperse liquid crystal-in-water emulsion
droplets. The droplet phase is the liquid crystal 4-pentyl-cyano-biphenyl (5CB). The drops have a
uniform radius in the range $b = 35-50\,\mu m$ and are immersed in an external water phase
containing Sodium Dodecyl Sulfate (SDS), which is at a concentration [5 to 20$\%$ weight/weight(w/w)]
much greater than its critical micellar concentration (CMC). The presence of surfactant at such high 
concentrations causes a spontaneous dissolution of the droplet  \cite{Peddireddy2012}, the result of 
which is sustained motion of the droplets. The direction of motion is due to a spontaneously broken 
symmetry \cite{Herminghaus2014}, which is dynamically sustained by the dissolution instability leading to an interfacial tension gradient 
around the droplet. As a result, the droplets translate through the aqueous background at speeds in 
the range of $v_s \sim 3-20\, \mu m /s$.

\subsection*{Active forces and torques}. The details of our dynamical model
for the active spheres, including long-ranged, many-body hydrodynamic
interactions in the presence of boundaries, are described below. The
spherical active droplet is modeled as an active sphere with a slip
velocity $\boldsymbol{v}_{i}^{\mathcal{A}}$ on its surface, as shown
in Eq.(1). We expand the slip as
\begin{alignat}{1}
\boldsymbol{v}_{i}^{\mathcal{A}}\big(\boldsymbol{R}_{i}+\boldsymbol{\rho}_{i}\big)=\sum_{l=1}^{\infty}\frac{1}{(l-1)!(2l-3)!!}\,\mathbf{V}_{i}^{(l)}\cdot\mathbf{Y}^{(l-1)}(\boldsymbol{\hat{\rho}}_{i}),\label{eq:boundaryFields-Yl}
\end{alignat}
in the irreducible basis of tensorial spherical harmonics $\mathbf{Y}^{(l)}(\boldsymbol{\hat{\rho}}_{i})=(-1)^{l}\rho_{i}^{l+1}\boldsymbol{\nabla}{}^{(l)}\rho_{i}^{-1},$
where $\boldsymbol{\nabla}^{(l)}=\boldsymbol{\nabla}_{\alpha_{1}}\dots\boldsymbol{\nabla}_{\alpha_{l}}$,
and $\boldsymbol{\rho}_{i}$ denotes the radius vector of $i$-th
particle. The expansion coefficients $\mathbf{V}_{i}^{(l)}$ are $l$-th
rank reducible Cartesian tensors. They can be written in terms of
three irreducible parts $\mathbf{V}_{i}^{(l\sigma)}$ of ranks $l,$
$l-1,$ and $l-2$ respectively, corresponding to the symmetric traceless
($\sigma=s$), antisymmetric ($\sigma=a$) and pure trace ($\sigma=t$)
parts. The first two modes of the slip, $\mathbf{V}_{i}^{(1s)}\equiv-\mathbf{V}_{i}^{\mathcal{A}}$
and $\mathbf{V}_{i}^{(2a)}\equiv-b\mathbf{\Omega}_{i}^{\mathcal{A}}$,
are respectively the active translational velocity and active angular
velocity for a sphere in an unbounded medium \cite{anderson1991,Stone1996,ghose2014irreducible}.
Explicitly, they are \begin{subequations}
\begin{alignat}{1}
\,\mathbf{V}_{i}^{\mathcal{A}}= & -\tfrac{1}{4\pi b^{2}}\int\boldsymbol{v}_{i}^{\mathcal{A}}(\boldsymbol{\rho}_{i})dS_{i},\label{eq:one-body}\\
\boldsymbol{\Omega}_{i}^{\mathcal{A}}= & -\tfrac{3}{8\pi b^{4}}\int\boldsymbol{\rho}_{i}\times\boldsymbol{v}_{i}^{\mathcal{A}}(\boldsymbol{\rho}_{i})dS_{i}.
\end{alignat}
\end{subequations}Given the slip, we seek expressions for the hydrodynamic
forces $\mathbf{F}_{i}^{H}$ and torques $\mathbf{T}_{i}^{H}$ on
the spheres. By linearity of Stokes flow, it is clear that these must
be of the form\begin{subequations}\label{force-formulation}
\begin{alignat}{1}
\mathbf{F}_{i}^{H}= & -\boldsymbol{\gamma}_{ij}^{TT}\mathbf{\cdot V}_{j}-\boldsymbol{\gamma}_{ij}^{TY}\mathbf{\cdot\,\boldsymbol{\Omega}}_{j}-\sum_{l\sigma=1s}^{\infty}\boldsymbol{\gamma}_{ij}^{(T,\,l\sigma)}\cdot\mathbf{V}_{j}^{(l\sigma)},\label{eq:linear-force-torque}\\
\mathbf{T}_{i}^{H}= & -\boldsymbol{\gamma}_{ij}^{RT}\mathbf{\cdot V}_{j}-\boldsymbol{\gamma}_{ij}^{RR}\mathbf{\cdot\,\boldsymbol{\Omega}}_{j}-\sum_{l\sigma=1s}^{\infty}\boldsymbol{\gamma}_{ij}^{(R,\,l\sigma)}\cdot\mathbf{V}_{j}^{(l\sigma)},\label{eq:torque-expression}
\end{alignat}
\end{subequations}where repeated particle indices are summed over
and Cartesian indices, implicit in the bold-face notation, are fully
contracted. The $\boldsymbol{\gamma}_{ij}^{\alpha\beta}$ with ($\alpha,\beta=T,R$)
are the usual Stokes friction tensors \cite{ladd1988,mazur1982,kim2005},
while the $\boldsymbol{\gamma}_{ij}^{(T,\,l\sigma)}$ and $\boldsymbol{\gamma}_{ij}^{(R,\,l\sigma)}$
are friction tensors associated with the slip \cite{singh2016crystallization,singh2016generalized}.
Explicit forms of these generalized friction tensors can be obtained
from the solution of the boundary integral equation of the Stokes
problem, in terms of Green's function satisfying the appropriate boundary
conditions \cite{singh2016crystallization,singh2016generalized}.
In general, the tensors are long-ranged, many-body functions of Stokes
flow and encode the dissipative hydrodynamic interactions between
the active spheres mediated by the intervening fluid. Brownian forces
on swimmers are typically $O(k_{B}T/b)\text{\ensuremath{\approx}}10^{-16}\,$N,
while the typical active forces on our swimmers are of the order $F_{\mathcal{A}}=6\pi\eta bv_{s}\text{\ensuremath{\approx}}10^{-11}\,\text{N}$.
Thus, the active hydrodynamic forces are orders of magnitude larger
than the Brownian forces, and we therefore ignore the effects of thermal
fluctuations in the fluid. The translational and rotational motions
of the spheres are then found by setting the net hydrodynamic forces
and torques to zero.

\subsection*{Rigid body motion of active colloids}. Using the above expressions
for the forces and torques in Newton's equations and inverting the
linear system for the velocities and angular velocities yields explicit
dynamical equations for the rates of change of positions and orientations:\begin{subequations}\label{eq:RBM}
\begin{alignat}{1}
\mathsf{\mathbf{V}}_{i} & =\boldsymbol{\mu}_{ij}^{TT}\cdot\mathbf{F}_{j}^{P}+\boldsymbol{\mu}_{ij}^{TR}\cdot\mathbf{T}_{j}^{P}+\sum_{l\sigma=2s}^{\infty}\boldsymbol{\pi}_{ij}^{(T,l\sigma)}\cdot\mathbf{\mathsf{\mathbf{V}}}_{j}^{(l\sigma)}+\mathsf{\mathbf{V}}_{i}^{\mathcal{A}},\\
\mathsf{\mathbf{\Omega}}_{i} & =\boldsymbol{\mu}_{ij}^{RT}\cdot\mathbf{F}_{j}^{P}+\boldsymbol{\mu}_{ij}^{RR}\cdot\mathbf{T}_{j}^{P}+\sum_{l\sigma=2s}^{\infty}\boldsymbol{\pi}_{ij}^{(R,\,l\sigma)}\cdot\mathbf{\mathsf{\mathbf{V}}}_{j}^{(l\sigma)}+\mathsf{\mathbf{\Omega}}_{i}^{\mathcal{A}}.
\end{alignat}
\end{subequations}These are the principal dynamical equations of
our model. Here, $\boldsymbol{\mu}_{ij}^{\alpha\beta}$ are the familiar
mobility matrices \cite{kim2005}, while the propulsion tensors $\boldsymbol{\pi}_{ij}^{(\alpha,\,l\sigma)}$,
first introduced in \cite{singh2015many}, relate the rigid body motion
to modes of the active slip. They are given in terms of the mobility
matrices and the generalized friction tensors \cite{singh2016generalized},\begin{subequations}
\begin{alignat}{1}
-\boldsymbol{\pi}_{ij}^{(\text{T},\,l\sigma)} & =\boldsymbol{\mu}_{ik}^{TT}\cdot\boldsymbol{\gamma}_{kj}^{(T,\,l\sigma)}+\boldsymbol{\mu}_{ik}^{TR}\cdot\boldsymbol{\gamma}_{kj}^{(R,\,l\sigma)},\\
-\boldsymbol{\pi}_{ij}^{(R,\,l\sigma)} & =\boldsymbol{\mu}_{ik}^{RT}\cdot\boldsymbol{\gamma}_{kj}^{(T,\,l\sigma)}+\boldsymbol{\mu}_{ik}^{RR}\cdot\boldsymbol{\gamma}_{kj}^{(R,\,l\sigma)}.
\end{alignat}
\end{subequations}The propulsion tensors inherit the long-ranged,
many-body character of the generalized friction tensors and play a
role analogous to the mobility matrices. A method of calculating them
directly, without requiring inversion of the generalized friction
tensors, has been provided in \cite{singh2015many}. The solution
is obtained in terms of a Green's function of Stokes equation.
\begin{figure*}
\includegraphics[width=0.95\textwidth]{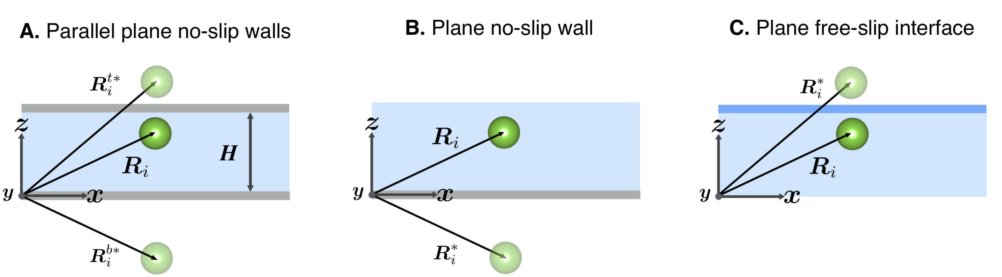}

\caption{The coordinate system used to describe the $i$-th spherical active
colloid and its image in following geometries of flow: \textbf{A}.
parallel no-slip plane walls, \textbf{B}. plane no-slip wall, and
\textbf{C}. plane free-slip surface. See the text for details \label{fig:6}}
\end{figure*}

We now list the Green's functions used for the various geometries
explored in the experiment.

\textit{(a) Plane no-slip wall}\textbf{: }We use the Lorentz-Blake
Green's function \cite{blake1971c} for swimmers near a no-slip wall.
The Lorentz-Blake tensor satisfies the no-slip condition at a plane
wall. We write it as
\begin{align}
G_{\alpha\beta}^{\text{w}}(\boldsymbol{R}_{i},\,\boldsymbol{R}_{j}) & =G_{\alpha\beta}^{\text{o}}(\boldsymbol{R}_{i}-\boldsymbol{R}_{j})+G_{\alpha\beta}^{*}(\boldsymbol{R}_{i},\,\boldsymbol{R}_{j}^{*}),\label{eq:lorentzBlake}
\end{align}
 where $\boldsymbol{G}^{\text{o}}(\boldsymbol{r})=(\nabla^{2}\boldsymbol{I}-\boldsymbol{\nabla}\boldsymbol{\nabla})\,\frac{r}{8\pi\eta}$
is the Oseen tensor and $\boldsymbol{G}^{*}(\boldsymbol{R}',\,\boldsymbol{R}^{*})$
is the correction, due to the image system, necessary to satisfy the
boundary condition at the no-slip wall. The expression for the correction
to Green's function near a plane wall is\textcolor{black}{
\begin{alignat}{1}
G_{\alpha\beta}^{*} & =\frac{1}{8\pi\eta}\Bigg[-\frac{\delta_{\alpha\beta}}{r^{*}}-\frac{r_{\alpha}^{*}r_{\beta}^{*}}{r^{*^{3}}}+2h^{2}\bigg(\frac{\delta_{\alpha\nu}}{r^{*^{3}}}-\frac{3r_{\alpha}^{*}r_{\nu}^{*}}{r^{*^{5}}}\bigg)\mathcal{M}_{\nu\beta}\nonumber \\
 & -2h\bigg(\frac{r_{3}^{*}\delta_{\alpha\nu}+\delta_{\nu3}r_{\alpha}^{*}-\delta_{\alpha3}r_{\nu}^{*}}{r^{*3}}-\frac{3r_{\alpha}^{*}r_{\nu}^{*}r_{3}^{*}}{r^{*^{5}}}\bigg)\mathcal{M}_{\nu\beta}\Bigg].
\end{alignat}
}The correction is obtained from the image of the swimmer with respect
to the wall located at $\boldsymbol{\boldsymbol{R}}^{*}=\boldsymbol{\mathcal{M}}\cdot\boldsymbol{\boldsymbol{R}}$,
where $\boldsymbol{\mathcal{M}}=\boldsymbol{I}-2\boldsymbol{\hat{z}}\boldsymbol{\hat{z}}$
is the mirror operator. Here $\boldsymbol{r}=\boldsymbol{\boldsymbol{R}}_{i}-\boldsymbol{\boldsymbol{R}}_{j}$,
$\boldsymbol{r}^{*}=\boldsymbol{\boldsymbol{R}}_{i}-\boldsymbol{\boldsymbol{R}}_{j}^{*}$,
and $h$ is the height of the swimmer from the wall. The coordinate
system is described in Fig. (\ref{fig:6}).

\textit{(b) Plane free-slip interface:}\textbf{ }We use the Green's
function obtained by Aderogba and Blake \cite{aderogba1976} for swimmers
near a free-slip air-water interface
\begin{alignat}{1}
G_{\alpha\beta}^{\text{s}}(\boldsymbol{R}_{i},\,\boldsymbol{R}_{j}) & =G_{\alpha\beta}^{\text{o}}(\boldsymbol{r})+(\delta_{\beta\alpha_{\rho}}\delta_{\alpha_{\rho}\gamma}-\delta_{\beta3}\delta_{3\gamma})G_{\alpha\gamma}^{0}(\boldsymbol{r}^{*}).
\end{alignat}

\textit{(c) Parallel plane no-slip walls:}\textbf{ }The Green's functions
for Stokes flow between parallel, planar no-slip walls has been obtained
by several authors. In the Hele-Shaw limit, where the separation between
the walls is of the order of the diameter of the swimmer, $H\sim2b$,
we use an approximate form of the Green's function \cite{liron1976stokes}
\begin{alignat}{1}
G_{\alpha\beta}^{\text{2w}}(\boldsymbol{R}_{i},\,\boldsymbol{R}_{j}) & =\frac{3hz(H-z)(H-h)}{\pi\eta H^{3}}\left(\frac{\delta_{\alpha\beta}}{2r_{_{\parallel}}^{2}}-\frac{r_{\alpha}r_{\beta}}{r_{_{\parallel}}^{4}}\right).\label{eq:2walls-G-}
\end{alignat}
Here $r_{_{\parallel}}$ is a measure of the distances in the plane
parallel to the walls.

At larger separations, $H\sim10b$, we use the solution of \cite{staben2003motion},
which employs a series sum of the Lorentz-Blake tensor. This is given
as
\begin{alignat}{1}
G_{\alpha\beta}^{\text{2w}}(\boldsymbol{R}_{i},\,\boldsymbol{R}_{j}) & =G_{\alpha\beta}^{\text{o}}(\boldsymbol{R}_{i},\,\boldsymbol{R}_{j})+G_{\alpha\beta}^{*}(\boldsymbol{R}_{i},\,\boldsymbol{R}_{j}^{b*})\nonumber \\
 & +G_{\alpha\beta}^{*}(\boldsymbol{R}_{i},\,\boldsymbol{R}_{j}^{t*})+G_{\alpha\beta}^{\delta}.\label{eq:2walls-G--1}
\end{alignat}
Here $\boldsymbol{R}_{j}^{b*}$ and $\boldsymbol{R}_{j}^{t*}$ are
the images of the swimmer at $\,\boldsymbol{R}_{j}$ due to the bottom
and top walls respectively. $G_{\alpha\beta}^{\delta}$ is the additional
contribution required to satisfy the boundary conditions at the two
walls. This is approximate by a series sum of the Lorentz-Blake tensor,
Eq. (\ref{eq:lorentzBlake}), about the two walls.

\subsection*{Estimation of the slip flow}. Given the slip, the fluid flow
is obtained exactly in terms of the boundary integral representation
of Stokes flow. Briefly, the flow is expressed as an infinite series
of boundary integrals, each integral corresponding to a mode of the
slip, and these integrals are evaluated exactly in terms of the Green's
function of Stokes flow, making use of its biharmonic property \cite{singh2015many,singh2016generalized,singh2017fluctuation}.
The experimental flow is well-approximated by a truncated version
of the slip expansion of Eq.(\ref{eq:boundaryFields-Yl}) which retains
the following three leading terms
\begin{alignat}{1}
\boldsymbol{v}_{i}^{\mathcal{A}}(\boldsymbol{\rho}_{i})= & -\mathbf{V}_{i}^{\mathcal{A}}+\tfrac{1}{15}\mathbf{V}_{i}^{(3t)}\cdot\mathbf{Y}^{(2)}(\boldsymbol{\rho}_{i})\nonumber \\
 & +\mathbf{V}_{i}^{(2s)}\cdot\mathbf{Y}^{(1)}(\boldsymbol{\rho}_{i})-\tfrac{1}{75}\mathbf{V}_{i}^{(4t)}\hspace{-0.12cm}\cdot\mathbf{Y}^{(3)}(\boldsymbol{\rho}_{i}).\label{eq:slipTruncation}
\end{alignat}
This includes contributions from $l\sigma=2s,\,3t$ and $4t$ modes
of the slip, which correspond respectively to flows with dipolar,
degenerate quadrupolar, and quadrupolar symmetry. \textcolor{black}{We
use uniaxial parametrization of the slip coefficients in terms of
the particle orientation $\boldsymbol{p}_{i}$ as: $\mathbf{V}_{i}^{\mathcal{A}}=v_{s}\mathbf{Y}^{(1)}(\boldsymbol{p}_{i})$
$\mathbf{V}_{i}^{(2s)}=V_{0}^{(2s)}\mathbf{Y}^{(2)}(\boldsymbol{p}_{i})$,
$\mathbf{V}_{i}^{(3t)}=V_{0}^{(3t)}\mathbf{Y}^{(1)}(\boldsymbol{p}_{i})$,
and $\mathbf{V}_{i}^{(4t)}=V_{0}^{(4t)}\mathbf{Y}^{(2)}(\boldsymbol{p}_{i})$
where $Y_{\alpha}^{(1)}(\boldsymbol{p})=p_{\alpha}$ and $Y_{\alpha\beta}^{(2)}(\boldsymbol{p})=p_{\alpha}p_{\beta}-\tfrac{1}{3}\delta_{\alpha\beta}$.
T}he strengths of the modes - $V_{0}^{(2s)}=0.05$, $V_{0}^{(3t)}=0.25,$
and $V_{0}^{(4t)}=0.10$ - are determined by fitting the experimental
and the theoretical fluid flows. The above form of the slip is then
used to obtain the flow and the forces and torques necessary to update
the positions and orientations of the swimmers in different geometries. 

\subsection*{Simulation Details.} The simulations are performed by numerical
integration of Eq. (\ref{eq:RBM}), where the series sum is truncated
based on the minimal model of the active slip in Eq. (\ref{eq:slipTruncation}).
The coefficients of slip expansion are chosen to ensure that the theoretical
flow field is qualitatively similar to the experimentally measured
flow field. These parameters are then kept fixed for all the simulations
reported. The evaluation of the rigid body motion of the colloids
and the flow disturbance created by them has been performed using
PyStokes \cite{pystokes}. The numerical integration of the resulting
equations are performed using an adaptive time step integrator, which
uses the backward differentiation formula \cite{langtangen2012tutorial}.
Random packing of hard-spheres \cite{skoge2006packing} is used as
the initial distribution of particles in all the simulations. The colloid-colloid
and the colloid-wall repulsive interaction is modeled using the short-ranged
repulsive WCA potential \cite{weeks1971role}, which is given as,
$U(r)=\epsilon\left(\frac{r_{min}}{r}\right)^{12}-2\epsilon\left(\frac{r_{min}}{r}\right)^{6}+\epsilon,$
for $r<r_{min}$ and zero otherwise, where $\epsilon$ is the potential
strength. The WCA parameters for particle-particle repulsion are:
$r_{min}=4.4,\,\epsilon=0.08$, while for the particle-wall repulsion
we choose $r_{min}=2.4,\,\epsilon=0.08$. The number of particles
$N$ used, for respective plots, are: Fig. (2A): $N=6$; Fig. (2B):
$N=900$; Fig. (2C), (2D), (4A), and (4B): $N=1024$. Other parameters
used in the simulations are: radius of particle ($b=1$), self-propulsion
speed $v_{s}=0.15$, the strength of the bottom-heaviness ($T_{0}=0.2$) and dynamic viscosity $\eta=0.1$.
The phase diagrams of Fig. (3) are obtained using $N=16$ particles
for panels (A) and (B), while $N=128$ particles have been used to
obtain panels (C) and (D). These parameters have been used in generating
all the figures except the phase diagrams of Fig. (3), where strengths
of the symmetric irreducible dipole and vector quadrupole have been
varied to map the phase diagram.


\begin{thebibliography}{10}

\bibitem{brennen1977}
Brennen, C \& Winet, H.
\newblock (1977) {Fluid mechanics of propulsion by cilia and flagella}.
\newblock {\em Annu. Rev. Fluid Mech.} {\bf 9}, 339--398.

\bibitem{ebbens2010pursuit}
Ebbens, S.~J \& Howse, J.~R.
\newblock (2010) In pursuit of propulsion at the nanoscale.
\newblock {\em Soft Matter} {\bf 6}, 726--738.

\bibitem{Thutupalli2011}
Thutupalli, S, Seemann, R,  \& Herminghaus, S.
\newblock (2011) Swarming behavior of simple model squirmers.
\newblock {\em New J. Phys.} {\bf 13}, 073021.

\bibitem{singh2016crystallization}
Singh, R \& Adhikari, R.
\newblock (2016) Universal hydrodynamic mechanisms for crystallization in
  active colloidal suspensions.
\newblock {\em Phys. Rev. Lett.} {\bf 117}, 228002.

\bibitem{palacci2013living}
Palacci, J, Sacanna, S, Steinberg, A.~P, Pine, D.~J,  \& Chaikin, P.~M.
\newblock (2013) Living crystals of light-activated colloidal surfers.
\newblock {\em Science} {\bf 339}, 936--940.

\bibitem{petroff2015fast}
Petroff, A.~P, Wu, X.-L,  \& Libchaber, A.
\newblock (2015) Fast-moving bacteria self-organize into active two-dimensional
  crystals of rotating cells.
\newblock {\em Phys. Rev. Lett.} {\bf 114}, 158102--158106.

\bibitem{Thutupalli2013}
Thutupalli, S \& Herminghaus, S.
\newblock (2013) Tuning active emulsion dynamics via surfactants and topology.
\newblock {\em Eur. Phys. J. E} {\bf 36}, 9905.

\bibitem{Herminghaus2014}
Herminghaus, S, Maass, C.~C, Kr\"{u}ger, C, Thutupalli, S, Goehring, L,  \&
  Bahr, C.
\newblock (2014) {Interfacial mechanisms in active emulsions.}
\newblock {\em Soft Matter} pp. 7008--7022.

\bibitem{Izri2014}
Izri, Z, Linden, M. N. V.~D,  \& Dauchot, O.
\newblock (2014) Self-propulsion of pure water droplets by spontaneous
  marangoni-stress-driven motion.
\newblock {\em Phys. Rev. Lett.} {\bf 248302}, 1--5.

\bibitem{Shani2014}
Shani, I, Beatus, T, Bar-Ziv, R.~H,  \& Tlusty, T.
\newblock (2014) {Long-range orientational order in two-dimensional
  microfluidic dipoles}.
\newblock {\em Nat. Phys.} {\bf 10}, 1--5.

\bibitem{Cavagna2014}
Cavagna, A \& Giardina, I.
\newblock (2014) Bird flocks as condensed matter.
\newblock {\em Annu. Rev. Condens. Matter Phys.} {\bf 5}, 183--207.

\bibitem{Sokolov2007}
Sokolov, A, Aranson, I, Kessler, J,  \& Goldstein, R.
\newblock (2007) Concentration dependence of the collective dynamics of
  swimming bacteria.
\newblock {\em Phys. Rev. Lett.} {\bf 98}, 1--4.

\bibitem{woodhouse2012}
Woodhouse, F.~G \& Goldstein, R.~E.
\newblock (2012) Spontaneous circulation of confined active suspensions.
\newblock {\em Phys. Rev. Lett.} {\bf 109}, 168105.

\bibitem{Wioland2013}
Wioland, H, Woodhouse, F.~G, Dunkel, J, Kessler, J.~O,  \& Goldstein, R.~E.
\newblock (2013) Confinement stabilizes a bacterial suspension into a spiral
  vortex.
\newblock {\em Phys. Rev. Lett.} {\bf 110}, 268102.

\bibitem{Bricard2015}
Bricard, A. et. al.
\newblock (2015) Emergent vortices in populations of colloidal rollers.
\newblock {\em Nat. Commun.} {\bf 6}, 7470.

\bibitem{Solon2015}
Solon, A.~P. et.~al.
\newblock (2015) Pressure is not a state function for generic active fluids.
\newblock {\em Nat. Phys.} {\bf 11}, 673.

\bibitem{Fily2014}
Y.~Fily, A.~Baskaran, M.~H.
\newblock (2014) Dynamics of self-propelled particles under strong confinement.
\newblock {\em Soft Matter} {\bf 10}, 5609--5617.

\bibitem{niu2017self}
Niu, R, Palberg, T,  \& Speck, T.
\newblock (2017) Self-assembly of colloidal molecules due to self-generated
  flow.
\newblock {\em Phys. Rev. Lett.} {\bf 119}, 028001.

\bibitem{Souslov2017}
Souslov, A, van Zuiden, B.~C, Bartolo, D,  \& Vitelli, V.
\newblock (2017) {Topological sound in active-liquid metamaterials}.
\newblock {\em Nat. Phys.}

\bibitem{Woodhouse2017}
Woodhouse, F.~G \& Dunkel, J.
\newblock (2017) {Active matter logic for autonomous microfluidics}.
\newblock {\em Nat. Commun.} {\bf 8}, 15169.

\bibitem{Prishchepa2005}
Prishchepa, O.~O, Shabanov, A.~V,  \& Zyryanov, V.~Y.
\newblock (2005) {Director configurations in nematic droplets with
  inhomogeneous boundary conditions}.
\newblock {\em Phys. Rev. E} {\bf 72}, 1--11.

\bibitem{ghose2014irreducible}
Ghose, S \& Adhikari, R.
\newblock (2014) Irreducible representations of oscillatory and swirling flows
  in active soft matter.
\newblock {\em Phys. Rev. Lett.} {\bf 112}, 118102.

\bibitem{singh2015many}
Singh, R, Ghose, S,  \& Adhikari, R.
\newblock (2015) Many-body microhydrodynamics of colloidal particles with
  active boundary layers.
\newblock {\em J. Stat. Mech} {\bf 2015}, P06017.

\bibitem{singh2016generalized}
Singh, R \& Adhikari, R.
\newblock (2018) Generalized {S}tokes laws for active colloids and their
  applications.
\newblock {\em J. Phys. Commun.} {\bf 2}, 025025.

\bibitem{lighthill1952}
Lighthill, J.~M.
\newblock (1952) {On the squirming motion of nearly spherical deformable bodies
  liquids at very small reynold number}.
\newblock {\em Comm. Pure Appl. Maths,(5)} {\bf 5}, 108--118.

\bibitem{blake1971a}
Blake, J.~R.
\newblock (1971) {A spherical envelope approach to ciliary propulsion}.
\newblock {\em J. Fluid Mech.} {\bf 46}, 199--208.

\bibitem{kondepudi2014modern}
Kondepudi, D \& Prigogine, I.
\newblock (1998) {\em Modern thermodynamics: from heat engines to dissipative
  structures}.
\newblock (John Wiley \& Sons).

\bibitem{touchette2009large}
Touchette, H.
\newblock (2009) The large deviation approach to statistical mechanics.
\newblock {\em Phys. Rep.} {\bf 478}, 1--69.

\bibitem{Drescher2009}
Drescher, K, Leptos, K.~C, Tuval, I, Ishikawa, T, Pedley, T.~J,  \& Goldstein,
  R.~E.
\newblock (2009) {Dancing volvox: Hydrodynamic bound states of swimming algae}.
\newblock {\em Phys. Rev. Lett.} {\bf 102}, 1--4.

\bibitem{Negi2005}
Negi, A.~S, Sengupta, K,  \& Sood, A.~K.
\newblock (2005) Frequency-dependent shape changes of colloidal clusters under
  transverse electric field.
\newblock {\em Langmuir} {\bf 21}, 11623--11627.

\bibitem{Sapozhnikov2003}
Sapozhnikov, M.~V \& et. al.
\newblock (2003) Dynamic self-assembly and patterns in electrostatically driven
  granular media.
\newblock {\em Phys. Rev. Lett.} {\bf 90}, 114301.

\bibitem{Schaller2010}
Schaller, V, Weber, C, Semmrich, C, Frey, E,  \& Bausch, A.~R.
\newblock (2010) {Polar patterns of driven filaments}.
\newblock {\em Nature} {\bf 467}, 73--77.

\bibitem{ohta2014traveling}
Ohta, T \& Yamanaka, S.
\newblock (2014) Traveling bands in self-propelled soft particles.
\newblock {\em Eur. Phys. J. ST} {\bf 223}, 1279--1291.

\bibitem{Solon2015a}
Solon, A.~P, Chat{\'e}, H,  \& Tailleur, J.
\newblock (2015) From phase to microphase separation in flocking models: The
  essential role of nonequilibrium fluctuations.
\newblock {\em Phys. Rev. Lett.} {\bf 114}, 068101.

\bibitem{cates2015}
Cates, M.~E \& Tailleur, J.
\newblock (2015) Motility-induced phase separation.
\newblock {\em Annu. Rev. Condens. Mat. Phys.} {\bf 6}, 219--244.

\bibitem{Peddireddy2012}
Peddireddy, K, Kumar, P, Thutupalli, S, Herminghaus, S,  \& Bahr, C.
\newblock (2012) {Solubilization of thermotropic liquid crystal compounds in
  aqueous surfactant solutions}.
\newblock {\em Langmuir} {\bf 28}, 12426--12431.

\bibitem{anderson1991}
Anderson, J \& Prieve, D.
\newblock (1991) Diffusiophoresis caused by gradients of strongly adsorbing
  solutes.
\newblock {\em Langmuir} {\bf 7}, 403--406.

\bibitem{Stone1996}
Stone, H.~A \& Samuel, A. D.~T.
\newblock (1996) Propulsion of microorganisms by surface distortions.
\newblock {\em Phys. Rev. Lett.} {\bf 77}, 4102--4104.

\bibitem{ladd1988}
Ladd, A.~J.~C.
\newblock (1988) Hydrodynamic interactions in a suspension of spherical
  particles.
\newblock {\em J. Chem. Phys.} {\bf 88}, 5051--5063.

\bibitem{mazur1982}
Mazur, P \& van Saarloos, W.
\newblock (1982) Many-sphere hydrodynamic interactions and mobilities in a
  suspension.
\newblock {\em Physica A: Stat. Mech. Appl.} {\bf 115}, 21--57.

\bibitem{kim2005}
Kim, S \& Karrila, S.~J.
\newblock (1992) {\em Microhydrodynamics: Principles and Selected
  Applications}.
\newblock (Butterworth-Heinemann).

\bibitem{blake1971c}
Blake, J.~R.
\newblock (1971) A note on the image system for a stokeslet in a no-slip
  boundary.
\newblock {\em Proc. Camb. Phil. Soc.} {\bf 70}, 303--310.

\bibitem{aderogba1976}
Aderogba, K.
\newblock (1976) On stokeslets in a two-fluid space.
\newblock {\em J. Eng. Math.} {\bf 10}, 143--151.

\bibitem{liron1976stokes}
Liron, N \& Mochon, S.
\newblock (1976) Stokes flow for a stokeslet between two parallel flat plates.
\newblock {\em J. Eng. Math.} {\bf 10}, 287--303.

\bibitem{staben2003motion}
Staben, M.~E, Zinchenko, A.~Z,  \& Davis, R.~H.
\newblock (2003) Motion of a particle between two parallel plane walls in
  low-{R}eynolds-number {P}oiseuille flow.
\newblock {\em Phys. Fluids} {\bf 15}, 1711--1733.

\bibitem{singh2017fluctuation}
Singh, R \& Adhikari, R.
\newblock (2017) Fluctuating hydrodynamics and the {B}rownian motion of an
  active colloid near a wall.
\newblock {\em Eur. J. Comp. Mech} {\bf 26}, 78--97.

\bibitem{pystokes}
Singh, R, Laskar, A,  \& Adhikari, R.
\newblock (2014) Pystokes: Hampi.

\bibitem{langtangen2012tutorial}
Langtangen, H.~P \& Wang, L.
\newblock (2012) Odespy.

\bibitem{skoge2006packing}
Skoge, M, Donev, A, Stillinger, F.~H,  \& Torquato, S.
\newblock (2006) Packing hyperspheres in high-dimensional euclidean spaces.
\newblock {\em Phys. Rev. E} {\bf 74}, 041127.

\bibitem{weeks1971role}
Weeks, J.~D, Chandler, D,  \& Andersen, H.~C.
\newblock (1971) Role of repulsive forces in determining the equilibrium
  structure of simple liquids.
\newblock {\em J. Chem. Phys.} {\bf 54}, 5237--5247.

\end{thebibliography}
\end{document}